\documentclass{sciart}
\pdfoutput=1

\DeclareMathOperator{\Median}{median}
\DeclareMathOperator{\Diag}{diag}
\newcommand*{\R}{\mathbb{R}}
\newcommand*{\mat}[1]{\bm{\mathrm{#1}}}

\begin{document}

\title{Large-scale Data Integration \\
using Matrix Denoising and Geometric Factor Matching}
\author[1,*]{Felix Held}
\affil[1]{Department of Mathematical Sciences, Chalmers University of Technology and University of Gothenburg, Gothenburg, Sweden}
\affil[*]{Correspondence to: Felix Held \textlangle felix.held@gmail.com\textrangle}
\date{}

\maketitle

\begin{abstract}
Unsupervised integrative analysis of multiple data sources has become common place and scalable algorithms are necessary to accommodate ever increasing availability of data. Only few currently methods have estimation speed as their focus, and those that do are only applicable to restricted data layouts such as different data types measured on the same observation units. We introduce a novel point of view on low-rank matrix integration phrased as a graph estimation problem which allows development of a method, \emph{large-scale Collective Matrix Factorization (lsCMF)}, which is able to integrate data in flexible layouts in a speedy fashion. It utilizes a matrix denoising framework for rank estimation and geometric properties of singular vectors to efficiently integrate data. The quick estimation speed of lsCMF while retaining good estimation of data structure is then demonstrated in simulation studies.

An implementation of the algorithm is available as a Python package at \url{https://github.com/cyianor/lscmf}.
\end{abstract}


\section{Introduction}

Integrative analysis of multiple data sources is a problem addressed, for example, in biostatistics \parencite[e.g.,][]{Singh2019,Almstedt2020,Argelaguet2021} to perform exploratory data analysis and find features explaining relevant variance in the data, or to be able to predict drug effect on biological targets in unseen scenarios. Integrating side information with a main data source has been found to improve prediction in collaborative filtering \parencite[e.g.,][]{SinghGordon2008,Su2009} with the goal of improving performance in recommender systems.

Data integration methods are often unsupervised and aim to leverage shared variance between data sources to improve either predictive performance of missing or unseen data, or, alternatively, to find interpretable factors that describe variance in the data within single data sources as well as between them.
Data integration of collections of two or more matrix data sources is a widely studied topic with some prominent examples being CCA \parencite{Hotelling1936}, IBFA \parencite{Tucker1958}, JIVE \parencite{Lock2013a}, MOFA/MOFA+ \parencite{Argelaguet2018,Argelaguet2020}, or GFA \parencite{Klami2015}.

Most data integration methods aim to model collections of data sources as signal plus noise. Integration is then facilitated by partitioning the variance present in the signal into subspaces shared with some or all other data sources and those specific to a single input.
Using low-rank models in a matrix factorization setting has proven to be effective in modeling relationships among data sources.
Formally, given a set of views \(\mathcal{V} = \{1, \dots, m\}\), which describe data modalities such as observation units and data types, their dimensions \(p_i\) for \(i \in \mathcal{V}\), and a data source layout \(\mathcal{I} = \{(i, j) \in \mathcal{V} \times \mathcal{V}: \mat{Y}_{ij} \in \R^{p_i \times p_j} \text{ is observed}\}\), data is modeled as
\begin{equation*}
\mat{Y}_{ij} = \mat{X}_{ij} + \mat{Z}_{ij}
\end{equation*}
where \(\mat{X}_{ij}\) is the low-rank signal and \(\mat{Z}_{ij}\) is a noise term. Commonly, some form of a matrix factorization model is used to represent the low-rank signal such as \(\mat{X}_{ij} = \mat{U}_i \mat{U}_j^\top\) \parencite[e.g.,][]{Klami2014,Kallus2019} where \(\mat{U}_i \in \R^{p_i \times k}\) for \(i \in \mathcal{V}\) and \(k \ll \min_i p_i\). Some data layouts that are supported by existing methods are shown in \cref{fig:graph-layouts}A-C.


Despite the ever-growing availability of data, a major limitation of existing methods is scalability. Matrix-based data integration typically results in the formulation of an optimization problem, either by explicit formulation \parencite[e.g.,][]{Lock2013a,Kallus2019,Gaynanova2019} or as a result of the application of variational inference (VI) to a Bayesian problem formulation \parencite[e.g.,][]{Klami2014,Klami2015,Argelaguet2018,Argelaguet2020}.
The drawback of optimization-based methods is that they typically do not scale well to large amounts of data. One possible exception is \textcite{Argelaguet2020} where mean-field VI was replaced with stochastic VI making it possible to apply their method on larger datasets. However, real computational benefits of this approach can only be gained when using specialized hardware such as GPUs which makes the method less accessible.
Some alternative approaches exist such as AJIVE \parencite{Feng2018}, a reformulation of JIVE, and D-CCA  \parencite{Shu2020}. Both of these methods first find low-rank approximations of input matrices \(\hat{\mat{Y}}_{ij}\) and then find components that describe globally shared variance among data sources through some variant of CCA. This strategy can be much faster than the optimization-based approach, assuming that the two steps are implemented in an efficient fashion. Note that D-CCA can only be applied to two matrices, however, a generalization exists \parencite[D-GCCA]{Shu2022b}, which, like AJIVE, generalizes to \(n\) matrices of different data types measured on the same group. However, AJIVE and D-GCCA only find globally shared or individual subspaces of variance among data sources. To the best of the author's knowledge there is no data integration method for flexible layouts such as any combination of data matrices in \cref{fig:graph-layouts} which is scalable.





\section{Contributions}

This paper introduces \emph{large-scale Collective Matrix Factorization (lsCMF)}, a scalable algorithm for data integration of flexible layouts of data matrices.
\begin{enumerate}
\item A novel reformulation of the low-rank data integration problem is proposed within the context of graph structures. (\cref{sec:graph-interpretation})
\item Leveraging a matrix denoising framework \parencite{Gavish2017} and geometric properties of singular vectors in low-rank models under additive noise \parencite{BenaychGeorges2012}, a factor matching algorithm is formulated which can partition flexible data collections into shared (globally and among subsets) or individual variance components. (\cref{sec:asymptotic-model,sec:lscmf})
\item We demonstrate the performance of lsCMF in simulation studies and compare its performance to established methods. (\cref{sec:evaluation})
\end{enumerate}

\section{Graphical interpretation of data integration}%
\label{sec:graph-interpretation}

\begin{figure}
\begin{adjustbox}{max width=\textwidth}
\begin{tikzpicture}
\node[circle,draw] (DAN1) {1};

\node[circle,draw] (DAN2) at ($(DAN1)+(-1cm,0cm)$) {2};
\node[circle,draw] (DAN3) at ($(DAN1)+(0cm,1cm)$) {3};
\node[circle,draw] (DAN4) at ($(DAN1)+(1cm,0cm)$) {4};

\draw (DAN1) -- (DAN2);
\draw (DAN1) -- (DAN3);
\draw (DAN1) -- (DAN4);

\node (LA) at ($(DAN1)+(-1.5cm,1.75cm)$) {\large\textbf{A}};

\node[rectangle,draw,minimum width=1cm, minimum height=1.5cm,fill=white] (DAR2) at ($(DAN1)+(0,-2.75cm)$) {};
\node[rectangle,draw,minimum width=1cm, minimum height=1.5cm,fill=white] (DAR1) at ($(DAR2)+(-1.1cm,0)$) {};
\node[rectangle,draw,minimum width=1cm, minimum height=1.5cm,fill=white] (DAR3) at ($(DAR2)+(1.1cm,0)$) {};

\node[left=0.05cm of DAR1] (DAV1) {1};
\node[above=0.05cm of DAR1] (DAV2) {2};
\node[above=0.05cm of DAR2] (DAV3) {3};
\node[above=0.05cm of DAR3] (DAV4) {4};

\node[circle,draw] (DBN1) at ($(DAN1)+(3cm,0cm)$) {1};
\node[circle,draw,right=of DBN1] (DBN2) {2};

\node[circle,draw] (DBN3) at ($0.5*(DBN1)+0.5*(DBN2)+(-1.5cm,1cm)$) {3};
\node[circle,draw] (DBN4) at ($0.5*(DBN1)+0.5*(DBN2)+(0cm,1cm)$) {4};
\node[circle,draw] (DBN5) at ($0.5*(DBN1)+0.5*(DBN2)+(1.5cm,1cm)$) {5};

\draw (DBN1) -- (DBN3);
\draw (DBN1) -- (DBN4);
\draw (DBN1) -- (DBN5);
\draw (DBN2) -- (DBN3);
\draw (DBN2) -- (DBN4);
\draw (DBN2) -- (DBN5);

\node (LB) at ($(LA)+(3cm,0cm)$) {\large\textbf{B}};

\node[rectangle,draw,minimum width=1cm, minimum height=1.5cm,fill=white] (DBR14) at ($0.5*(DBN1)+0.5*(DBN2)+(0cm,-2cm)$) {};
\node[rectangle,draw,minimum width=1cm, minimum height=1.5cm,fill=white] (DBR13) at ($(DBR14)+(-1.1cm,0cm)$) {};
\node[rectangle,draw,minimum width=1cm, minimum height=1.5cm,fill=white] (DBR15) at ($(DBR14)+(1.1cm,0)$) {};

\node[rectangle,draw,minimum width=1cm, minimum height=1.5cm,fill=white] (DBR23) at ($(DBR13)+(0cm,-1.6cm)$) {};
\node[rectangle,draw,minimum width=1cm, minimum height=1.5cm,fill=white] (DBR24) at ($(DBR23)+(1.1cm,0)$) {};
\node[rectangle,draw,minimum width=1cm, minimum height=1.5cm,fill=white] (DBR25) at ($(DBR24)+(1.1cm,0)$) {};

\node[left=0.05cm of DBR13] (DBV1) {1};
\node[left=0.05cm of DBR23] (DBV2) {2};
\node[above=0.05cm of DBR13] (DBV3) {3};
\node[above=0.05cm of DBR14] (DBV4) {4};
\node[above=0.05cm of DBR15] (DBV5) {5};

\node[circle,draw] (DCN1) at ($(DBN2)+(3cm,0)$) {1};
\node[circle,draw] (DCN2) at ($(DCN1)+(-1cm,1cm)$) {2};
\node[circle,draw] (DCN3) at ($(DCN1)+(1cm,1cm)$) {3};

\draw (DCN1) -- (DCN2);
\draw (DCN1) -- (DCN3);
\draw (DCN2) -- (DCN3);

\node (LC) at ($(LB)+(4.5cm,0cm)$) {\large\textbf{C}};

\node[rectangle, draw, minimum width=1cm, minimum height=1.5cm, fill=white]
    (DCR1) at ($(DCN1)+(-0.25cm,-2.75cm)$) {};
\begin{scope}[shift={($(DCR1) + (0.5cm,0cm)$)}, canvas is yz plane at x=0]
    \node[rectangle, draw, minimum width=1.5cm, minimum height=1cm, fill=white, transform shape] (DCR2) at (-0.05, -0.75) {};
    
\end{scope}
\begin{scope}[shift={($(DCR1) + (0cm,0.75cm)$)}, canvas is xz plane at y=0]
    \node[rectangle, draw, minimum width=1cm, minimum height=1cm, fill=white, transform shape] (DCR3) at (-0.1, -0.75) {};
\end{scope}

\node[left=0.05cm of DCR1] (DCV1) {1};
\node[below=0.05cm of DCR1] (DCV21) {2};
\node (DCV31) at ($(DCR2) + (0.05cm,-1cm)$) {3};
\node (DCV22) at ($(DCR3) + (0.1cm,0.45cm)$) {2};
\node (DCV32) at ($(DCR3) + (-0.75cm,0.05cm)$) {3};

\node[circle,draw] (DDN1) at ($(DCN1)+(2.5cm,1.5cm)$) {1};
\node[circle,draw] (DDN2) at ($(DDN1)+(0cm,-2cm)$) {2};

\draw (DDN1) -- (DDN2);
\draw (DDN1) to[out=-115, in=115] (DDN2);
\draw (DDN1) to[out=-65, in=65] (DDN2);

\node (LD) at ($(LC)+(3.375cm,0cm)$) {\large\textbf{D}};

\path
    let
        \p1 = (DCR1),
        \p2 = (DDN2)
    in
        coordinate (DDR3orig) at (\x2+0.1cm, \y1+0.2cm);
\node[rectangle,draw,minimum width=1cm, minimum height=1.5cm,fill=white] (DDR3) at (DDR3orig) {};
\node[rectangle,draw,minimum width=1cm, minimum height=1.5cm,fill=white] (DDR2) at ($(DDR3)+(-0.1cm,-0.1cm)$) {};
\node[rectangle,draw,minimum width=1cm, minimum height=1.5cm,fill=white] (DDR1) at ($(DDR2)+(-0.1cm,-0.1cm)$) {};

\node[left=0.05cm of DDR1] (DDV1) {1};
\node[below=0.05cm of DDR1] (DDV2) {2};
\end{tikzpicture}
\end{adjustbox}
\caption{Illustrations of the identification of matrix integration layouts with graph structures. Nodes in the graphs represent views of the data and the presence of edges indicates that a data matrix describing the relationship between two views is present. Multiple matrices describing the same relationship can be present and are represented as multi-edges. \textbf{A} Multi-view layout \textbf{B} Grid layout \textbf{C} Augmented layout \textbf{D} Tensor-like layout}
\label{fig:graph-layouts}
\end{figure}
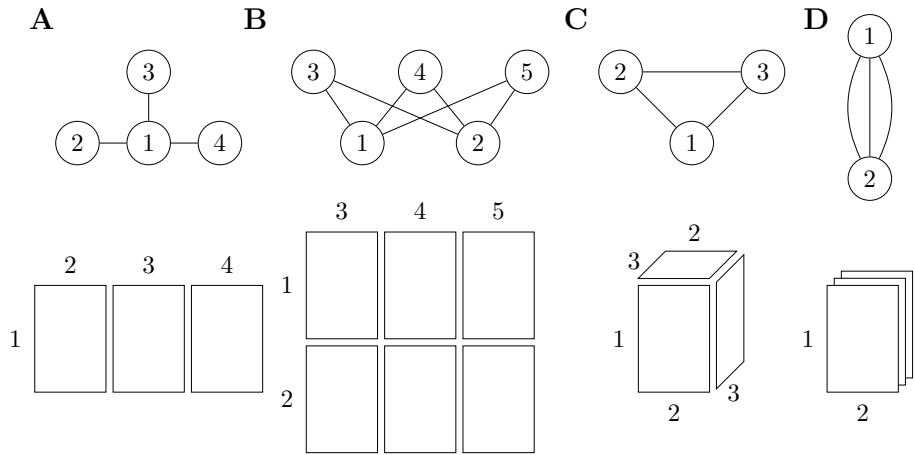

The general problem of data integration can be translated into graph structures that make it possible to reason about the problem in new ways. For an introduction to graph theory see e.g., \textcite{Diestel2017}. A graph \(G = (V, E)\) consists of a set of nodes or vertices \(V\) and a set of edges \(E\), describing connections between nodes. We consider \emph{undirected multi-graphs} as well as \emph{undirected hypergraphs}. In undirected graphs, edges are specified as sets between vertices, i.e., \(e = \{v_1, v_2\} \in E\) for \(v_1, v_2 \in V\). An undirected multi-graph allows that more than one edge may exist between the same pair of vertices. An undirected hypergraph considers edges \(e \in E\) to be non-empty subsets of \(V\). Edges in hypergraphs are called \emph{hyperedges}.

Typical matrix layouts considered in data integration are illustrated in \cref{fig:graph-layouts}.
The common \emph{multi-view layouts} (\cref{fig:graph-layouts}A) correspond to a single central node (samples) connected with single edges to a number of other nodes (data types).
\emph{Grid layouts} (\cref{fig:graph-layouts}B) are used in situations where the same data types were observed on two or more groups of samples. These layouts essentially result in a bipartite graph (a graph with two sets of nodes and edges only in-between sets) with groups on one side and data types on the other.
\emph{Augmented layouts} (\cref{fig:graph-layouts}C) introduce relationships between data types or sample groups and result in circular connections in the graph.
\emph{Tensor-like layouts} (\cref{fig:graph-layouts}D) describe repeated measurement scenarios or when the same experiment is repeated under modified conditions. They require multi-graphs to be represented and each edge stands for one layer in the tensor. Note that there is no data type or sample group connected with the third tensor dimension which is why we consider this a tensor-\emph{like} layout.

For simplicity, the graphs in \cref{fig:graph-layouts} will be called \emph{view graphs} since they describe relationships between views. Each data matrix is associated with its respective edge in the graph.

\definecolor{cred}{rgb}{0.95429001, 0.47795052, 0.19639176}
\definecolor{cgreen}{rgb}{0.20125317, 0.69079208, 0.47966761}
\definecolor{cblue}{rgb}{0.23299121, 0.63958655, 0.92607061}
\definecolor{cpurple}{rgb}{0.3984375 0.1992188 0.5976562}

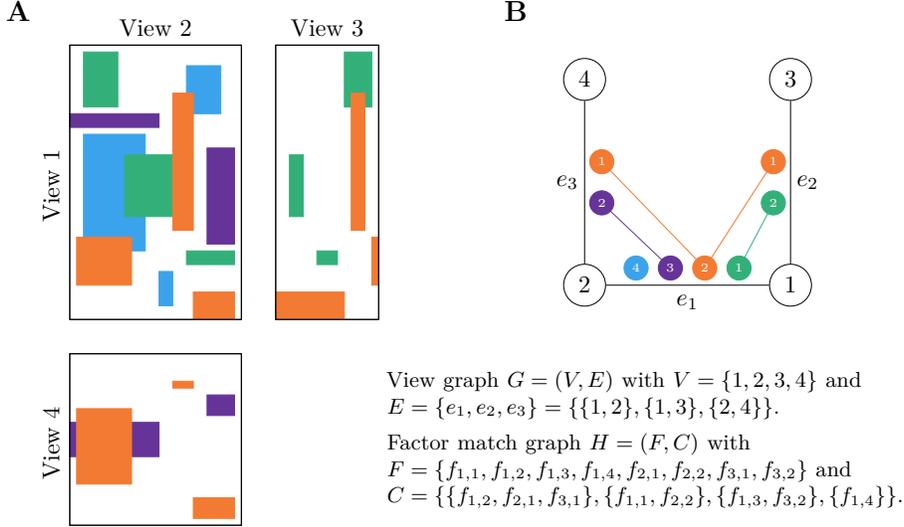
\begin{figure}[ht]
\centering
\begin{adjustbox}{max width=\textwidth}
\begin{tikzpicture}[factornode/.style={
    circle,
    white,
    inner sep=0pt,
    outer sep=0pt,
}]


\node at (-.75,4.5) {\large\textbf{A}};
\node at (6.5,4.5) {\large\textbf{B}};

\filldraw[color=cblue] (.2,1) rectangle ++(.9,1.7);
\filldraw[color=cblue] (1.7,3) rectangle ++(.5,.7);
\filldraw[color=cblue] (1.3,0.2) rectangle ++(.2,0.5);

\filldraw[color=cgreen] (.8,1.5) rectangle ++(.7,.9);
\filldraw[color=cgreen] (.2,3.1) rectangle ++(.5,.8);
\filldraw[color=cgreen] (1.7,0.8) rectangle ++(.7,.2);

\filldraw[color=cgreen] (3.2,1.5) rectangle ++(.2,.9);
\filldraw[color=cgreen] (4,3.1) rectangle ++(.4,.8);
\filldraw[color=cgreen] (3.6,0.8) rectangle ++(.3,.2);

\filldraw[color=cpurple] (0,2.8) rectangle ++(1.3,.2);
\filldraw[color=cpurple] (2,1.1) rectangle ++(.4,1.4);

\filldraw[color=cpurple] (0,-2) rectangle ++(1.3,.5);
\filldraw[color=cpurple] (2,-1.4) rectangle ++(.4,.3);

\filldraw[color=cred] (.1,.5) rectangle ++(.8,0.7);
\filldraw[color=cred] (1.5, 1.3) rectangle ++(.3,2);
\filldraw[color=cred] (1.8,0) rectangle ++(.6,.4);

\filldraw[color=cred] (4.4,.5) rectangle ++(.1,0.7);
\filldraw[color=cred] (4.1, 1.3) rectangle ++(.2,2);
\filldraw[color=cred] (3,0) rectangle ++(1,.4);

\filldraw[color=cred] (.1,-2.4) rectangle ++(.8,1.1);
\filldraw[color=cred] (1.5, -1) rectangle ++(.3,.1);
\filldraw[color=cred] (1.8,-2.9) rectangle ++(.6,.3);

\draw (0,0) rectangle ++(2.5,4) node[label={[shift={(-1.75ex,4ex)},rotate=90]left:View 1},label={above:View 2},pos=.5,minimum width=2.5cm, minimum height=4cm,rectangle,draw] {};

\draw (0,-3) rectangle ++(2.5,2.5) node[label={[shift={(-1.75ex,4ex)},rotate=90]left:View 4}, pos=.5,minimum width=2.5cm, minimum height=2.5cm,rectangle,draw] {};

\draw (3,0) rectangle ++(1.5,4) node[label={above:View 3},pos=.5,minimum width=1.5cm, minimum height=4cm,rectangle,draw] {};

\node[circle,draw] (DAN1) at (10.5, 0.5) {1};
\node[circle,draw] (DAN2) at ($(DAN1)+(-3cm,0cm)$) {2};
\node[circle,draw] (DAN3) at ($(DAN1)+(0cm,3cm)$) {3};
\node[circle,draw] (DAN4) at ($(DAN2)+(0cm,3cm)$) {4};

\filldraw[color=cgreen] ($(DAN1)+(-0.75cm, 0.25cm)$) circle (5pt) node[factornode] (f12-1) {\tiny 1};
\filldraw[color=cred] ($(DAN1)+(-1.25cm, 0.25cm)$) circle (5pt) node[factornode] (f12-2) {\tiny 2};
\filldraw[color=cpurple] ($(DAN1)+(-1.75cm, 0.25cm)$) circle (5pt) node[factornode] (f12-3) {\tiny 3};
\filldraw[color=cblue] ($(DAN1)+(-2.25cm, 0.25cm)$) circle (5pt) node[factornode] {\tiny 4};

\filldraw[color=cred] ($(DAN1)+(-0.25cm, 1.8cm)$) circle (5pt) node[factornode] (f13-1) {\tiny 1};
\filldraw[color=cgreen] ($(DAN1)+(-0.25cm, 1.2cm)$) circle (5pt) node[factornode] (f13-2) {\tiny 2};

\filldraw[color=cred] ($(DAN2)+(0.25cm, 1.8cm)$) circle (5pt) node[factornode] (f24-1) {\tiny 1};
\filldraw[color=cpurple] ($(DAN2)+(0.25cm, 1.2cm)$) circle (5pt) node[factornode] (f24-2) {\tiny 2};

\draw[color=cred] (f13-1) -- (f12-2) -- (f24-1);
\draw[color=cgreen] (f13-2) -- (f12-1);
\draw[color=cpurple] (f12-3) -- (f24-2);

\draw (DAN1) -- (DAN2) node[midway,yshift=-0.25cm] {\(e_1\)};
\draw (DAN1) -- (DAN3) node[midway,xshift=0.25cm] {\(e_2\)};
\draw (DAN2) -- (DAN4) node[midway,xshift=-0.25cm] {\(e_3\)};

\node[text width=7.5cm, anchor=west] at (4.5, -1.75) {\begin{minipage}[t]{7.75cm}\raggedright\small View graph \(G = (V, E)\) with \(V = \{1, 2, 3, 4\}\) and \(E = \{e_1, e_2, e_3\} = \{\{1, 2\}, \{1, 3\}, \{2, 4\}\}\).\\[1ex]

Factor match graph \(H = (F, C)\) with \(F = \{f_{1,1}, f_{1,2}, f_{1,3}, f_{1,4}, f_{2,1}, f_{2,2}, f_{3,1}, f_{3,2}\}\) and \(C = \{\{f_{1,2}, f_{2,1}, f_{3,1}\}, \{f_{1,1}, f_{2,2}\}, \{f_{1,3}, f_{3,2}\}, \{f_{1,4}\}\}.\)\end{minipage}};
\end{tikzpicture}
\end{adjustbox}
\caption{Example of an L-shaped layout of data sources containing low-rank signal which is shown as a sketch in \textbf{A}.
Colors indicate different components of the low-rank signals. Matching colors in different data sources indicate that these signals are described by a shared factor.
There is globally shared signal (\textcolor{cred}{orange}), signal shared between matrices \(\mat{X}_{12}\) and \(\mat{X}_{13}\) (\textcolor{cgreen}{green}), signal shared between matrices \(\mat{X}_{12}\) and \(\mat{X}_{24}\) (\textcolor{cpurple}{purple}), as well as individual signal in matrix \(\mat{X}_{12}\) (\textcolor{cblue}{blue}).
In \textbf{B} an illustration is shown of the view graph corresponding to the data layout (nodes and edges in black) as well as of the factor match graph (colored nodes and edges) corresponding to the component structure. Colors of hypergraph nodes and hyperedges match the colors of the corresponding components in \textbf{A}.}
\label{fig:hypergraph-equiv}
\end{figure}

Conceptually, factor directions are associated with nodes in the view graph and edges, associated with data sources, contain information about which factors are active in the data source. By introducing a hypergraph \(H = (F, C)\) which associates nodes for each active factor in an edge of the view graph, hyperedges can then be used to describe relationships between components. Consider the example illustrated in \cref{fig:hypergraph-equiv}A. An L-shaped layout of three matrices is presented with matrices containing low-rank signals consisting of different components. The view graph associated with this example is shown in \cref{fig:hypergraph-equiv}B. Focus now on edge \(e_1 = \{1, 2\}\) in the view graph. It corresponds to matrix \(\mat{X}_{12}\) which contains four different low-rank components. \(e_1\) is therefore associated with nodes \(f_{1,1}, f_{1,2}, f_{1,3}, f_{1,4} \in F\). Similarly for \(e_2\) and \(e_3\) all nodes in \(H\) can be determined. Edges in \(H\) are then chosen in such a way that they describe sharing of components, and therefore information, among matrices in \cref{fig:hypergraph-equiv}A. As an example, \(c_1 = \{f_{1,2}, f_{2,1}, f_{3,1}\}\) describes the globally shared component and \(c_4 = \{f_{1,4}\}\) describes the individual component in \(\mat{X}_{12}\).
The hypergraph corresponding to the view graph is illustrated in \cref{fig:hypergraph-equiv}B.
Since \(H\) describes a matching among factors, we call it a \emph{factor match graph}.

Note that hyperedges in a factor match graph are not arbitrary. In the example shown above, the hyperedge \(\{f_{2,1}, f_{3, 1}\}\) would not be allowed. A factor node associated with edge \(e_1\) is necessary as a bridging connection. Therefore, connections can only be established if two edges in the view graph share a node. Hyperedges can then propagate along the view graph as is the case for the globally shared component in the example above.

Formulating data integration models in this fashion is equivalent to a matrix low-rank formulation such as \(\mat{X}_{ij} = \mat{V}_i \Diag(x_1, \dots, x_r) \mat{V}_j^\top\) in the sense that factor vectors in \(\mat{V}_i\) are associated with the node \(v_i\) in the view graph, and non-zero singular values among \(x_1, \dots, x_r)\) can be associated with nodes in the factor match graph. However, this new formulation allows to think about the estimation problem in a fresh way.


In the following, we propose \emph{large-scale Collective Matrix Factorization (lsCMF)}, a methodology to estimate the factor match graphs such as described above.
The approach first constructs a factor match graph for each view separately, which essentially reduces the integration problem to a sequence of multi-view problems. The separate factor match graphs are then merged to produce the final integration estimate. We propose a scalable and fast algorithm to estimate view-specific factor match graphs, based on matrix denoising and geometric factor matching in an asymptotic regime, and show how factor match graphs can be merged efficiently. In \cref{sec:asymptotic-model}, some background is given on the denoising, model assumptions, and asymptotic results on angles between true signal singular vectors and their empirical counterparts. In \cref{sec:lscmf}, the core material of this paper is presented. A novel estimation process for factor factor match graphs based on geometric factor matching and the merging of factor match graphs. In \cref{sec:evaluation}, we demonstrate the performance of our method through simulation studies.


\section{An Asymptotic Model for Matrix Denoising}%
\label{sec:asymptotic-model}

The asymptotic framework introduced in \textcite{BenaychGeorges2012} will be used.
Assume that
\begin{equation}\label{eq:asymp-model}
\mat{Y}_k = \mat{X}_k + \frac{1}{\sqrt{n_k}} \mat{Z}_k
\end{equation}
where \(\mat{Y}_k, \mat{X}_k, \mat{Z}_k \in \R^{m_k \times n_k}\). The size of
the model is assumed to grow with \(k\) such that
\(m_k, n_k \rightarrow \infty\) for \(k \rightarrow \infty\).

It is assumed that \(\mat{X}_k\) is a deterministic matrix of fixed,
low rank \(r\). The matrix \(\mat{Z}_k\) is assumed to be random and
can be considered as observation noise. The only observed quantity is
\(\mat{Y}_k\). The goal of the analysis is to recover an estimate \(\hat{\mat{X}}_k\) 
of \(\mat{X}_k\) as best as possible.

A typical approach to try to recover \(\mat{X}_k\) is to find the best
rank \(r\) approximation of the observed data matrix \(\mat{Y}_k\).
This approximation is given by the Eckhard-Young-Mirsky theorem
\parencite{Eckart1936,Mirsky1960} and is the SVD of \(\mat{Y}_k\)
truncated after \(r\) terms. However, as noted in \textcite{Nadakuditi2011},
this forms an approximation to \emph{low-rank signal plus noise} instead
of the signal directly.

One approach to approximate the signal in a minimax optimal way, is the denoising
approach in \textcite{Gavish2017}. Their framework requires the following assumptions.


\subsection{Model Assumptions}%
\label{ssec:model-assumptions}

\begin{assumption}[Entries of \(\mat{Z}_k\)]\label{ass:entries-z}
The entries of \(\mat{Z}_k\) are \emph{iid} and have zero mean, unit
variance and finite fourth moment. In addition, we assume that the distribution of
the elements of \(\mat{Z}_k\) is orthogonally invariant\footnote{A common example is
\emph{iid} standard normal distributed noise.}. This means that for
orthogonal matrices \(\bm{A} \in \R^{m_k \times m_k}\) and
\(\bm{B} \in \R^{n_k \times n_k}\) the distribution of \(\bm{A} \bm{Z}_k \bm{B}\)
is the same as the distribution of \(\bm{Z}_k\).
\end{assumption}
\begin{assumption}[Signal singular values]\label{ass:signal-sing-vals}
It is assumed there is a fixed set of singular values
\((\bm{x}^{(1)}, \dots, \bm{x}^{(r)})\) where \(r > 0 \) is fixed (i.e.\@ does
not depend on \(k\)) and that there are matrices
\(\mat{U}_k \in \R^{m_k \times r}\) and \(\mat{V}_k \in \R^{n_k \times r}\)
with orthogonal columns such that
\begin{equation*}
\bm{X}_k = \mat{U}_k \Diag(\bm{x}^{(1)}, \dots, \bm{x}^{(r)}) \mat{V}_k^\top
\end{equation*}
\end{assumption}
\begin{assumption}[Asymptotic ratio]\label{ass:asymp-ratio}
It holds that \(m_k / n_k \rightarrow \beta \in \interval[open left]{0}{1}\) for
\(k \rightarrow \infty\).
\end{assumption}


\subsection{Optimal shrinkage}

The quality of the estimate \(\hat{\mat{X}}_k\) is measured with
Frobenius loss, i.e.\@ \(\norm{\mat{X}_k - \hat{\mat{X}}_k}_F\).
The loss cannot be evaluated in practice, but \textcite{Gavish2017}
showed that using singular value shrinkage on \(\mat{Y}_k\) achieves minimax
optimality in terms of asymptotic Frobenius loss for the model in
\cref{eq:asymp-model}.
The shrinkage function \parencite[Eq. 7]{Gavish2017} is
\begin{equation}\label{eq:frobenius-shrinkage-opt}
\eta(y) =
\begin{cases}
  \frac{\sqrt{(y^2 - \beta - 1)^2 - 4\beta}}{y} & y \geq 1 + \sqrt{\beta} \\
  0 & y < 1 + \sqrt{\beta} \\
\end{cases}
\end{equation}
where \(m_k / n_k \rightarrow \beta \in \interval[open left]{0}{1}\).
Let the SVD of \(\mat{Y}_k\) be given by
\begin{equation*}
\mat{Y}_k = \sum_{l = 1}^{m_k} \bm{y}_k^{(l)}
\mat{A}_k^{(:, l)} {\mat{B}_k^{(:, l)}}^\top,
\end{equation*}
where \(\bm{y}^{(1)} \geq \cdots \geq \bm{y}^{(m_k)} \geq 0\).
The denoised estimate \(\hat{\mat{X}}_k\) is then given by
\begin{equation}\label{eq:denoised-signal}
\hat{\mat{X}}_k =  \sum_{l = 1}^{m_k} \eta(\bm{y}_k^{(l)})
\mat{A}_k^{(:, l)} {\mat{B}_k^{(:, l)}}^\top.
\end{equation}


\subsection{Estimating an Unknown Noise Variance}

In practice, it is more likely to encounter matrices of the form
\begin{equation}\label{eq:asymp-model-noise}
\mat{Y}_k = \mat{X}_k + \sigma \mat{Z}_k
\end{equation}
with an unknown noise variance \(\sigma^2\). To be able to scale the data and work
in the model of \cref{eq:asymp-model} the noise variance needs to be estimated.
Assume that \Cref{ass:entries-z,ass:signal-sing-vals,ass:asymp-ratio} are fulfilled.
Denote the vector of singular values of \(\mat{Y}_k\) by \(\bm{y}_k\) and set
\(\bar{y}_k = \Median_l(\bm{y}_k^{(l)})\). Then \textcite{Gavish2014}
suggested the estimator
\begin{equation}\label{eq:sd-estimator}
\hat{\sigma}_k = \frac{\bar{y}_k}{\sqrt{n_k q_{\mathrm{MP}, \beta_k}}}
\end{equation}
where \(q_{\mathrm{MP}, \beta_k}\) is the median of the Mar{\v c}enko-Pastur
distribution with \(\sigma = 1\) and \(\beta_k = m_k / n_k\).
In particular, \parencite[Lemma 1]{Gavish2014} show that
\(\lim_{k \rightarrow \infty} \hat{\sigma}_k \overset{a.s.}{=} \sigma\).


\subsection{Asymptotic geometry of singular vectors}%
\label{ssec:asymptotic-geometry-singular-vectors}

Assume the true signal \(\mat{X}_k\) is of the form
\begin{equation*}
\mat{X}_k =
\sum_{l = 1}^r \bm{x}^{(l)} \mat{U}_k^{(:, l)} {\mat{V}_k^{(:, l)}}^\top
\end{equation*}
and assume in addition that \(\bm{x}^{(i)} \neq \bm{x}^{(j)}\) for all
\(1 \leq i < j \leq r\), which is reasonable in practice, and let the SVD of
\(\bm{Y}_k\) be
\begin{equation*}
\mat{Y}_k = \sum_{l = 1}^{m_k} \bm{y}_k^{(l)}
\mat{A}_k^{(:, l)} {\mat{B}_k^{(:, l)}}^\top.
\end{equation*}
Lemmas 2 and 3 in \textcite{Gavish2014} then show the following result.
\begin{lemma}\label{lmm:asymp-angles-values}
For \(1 \leq i \leq r\)
\begin{equation}\label{eq:asymp-singular-values}
\lim_{k \rightarrow \infty} \bm{y}_k^{(i)} \overset{a.s.}{=}
\begin{cases}
  \sqrt{%
    \big(\bm{x}^{(i)} + \frac{1}{\bm{x}^{(i)}}\big)
    \big(\bm{x}^{(i)} + \frac{\beta}{\bm{x}^{(i)}}\big)} &
  \bm{x}^{(i)} > \beta^{1/4} \\
  1 + \sqrt{\beta} &
  \bm{x}^{(i)} \leq \beta^{1/4}
\end{cases}
\end{equation}

For \(1 \leq i \leq j \leq r\) and \(\bm{x}^{(i)} > \beta^{1/4}\) it holds
that
\begin{equation}\label{eq:asymp-left-angles}
\lim_{k \rightarrow \infty}
\abs{\langle \mat{A}_k^{(:, i)}, \mat{U}_k^{(:, j)} \rangle} \overset{a.s.}{=}
\begin{cases}
\sqrt{\frac{%
{\bm{x}^{(i)}}^4 - \beta}{%
{\bm{x}^{(i)}}^4 + \beta {\bm{x}^{(i)}}^2}} & i = j \\
0 & i \neq j
\end{cases}
\end{equation}
\begin{equation}\label{eq:asymp-right-angles}
\lim_{k \rightarrow \infty}
\abs{\langle \mat{B}_k^{(:, i)}, \mat{V}_k^{(:, j)} \rangle} \overset{a.s.}{=}
\begin{cases}
  \sqrt{\frac{%
  {\bm{x}^{(i)}}^4 - \beta}{%
  {\bm{x}^{(i)}}^4 + {\bm{x}^{(i)}}^2}} & i = j \\
  0 & i \neq j
\end{cases}
\end{equation}
If \(\bm{x}^{(i)} \leq \beta^{1/4}\) then
\(\lim_{k \rightarrow \infty}
\langle \mat{A}_k^{(:, i)}, \mat{U}_k^{(:, j)} \rangle
\overset{a.s.}{=}
0
\overset{a.s.}{=}
\lim_{k \rightarrow \infty}
\langle \mat{B}_k^{(:, i)}, \mat{V}_k^{(:, j)} \rangle
\).
\end{lemma}
Illustrations of limiting behaviour are shown in \cref{fig:asymp-singular-values,fig:asymp-cos-of-angles}. \cref{lmm:asymp-angles-values} will play a crucial role in our description of geometric factor matching. Since the columns of \(\mat{A}_k\) and \(\mat{U}_k\) are normalized, the scalar products in \cref{eq:asymp-left-angles,eq:asymp-right-angles} describe absolute values of cosines of angles. These angles can help to determine whether or not two factors in two data matrices derive from the same underlying factor or not.


\subsection{High-dimensional rotations}%
\label{ssec:high-dim-rotations}

Rotations in two dimensions are typically thought of as rotating the plane around a point in 2D-space. Similarly, rotations in three dimensions can be seen as rotating 3D-space around an axis. In both cases, rotation only occurs in a two-dimensional space. In two dimensions, a zero-dimensional subspace remains fixed whereas in three dimensions a one-dimensional subspace remains fixed. Following this logic, simple rotations in \(n\) dimensions keep an \((n-2)\)-dimensional subspace fixed and rotate points parallel to a two-dimensional subspace \parencite{Aguilera2004}. However, more complex rotations can be built up from simple rotations. In even dimensions, it is possible that no subspace remains fixed, whereas there is always a fixed axis when rotation is performed in a space with odd dimension. This follows from the fact that rotation matrices have determinant 1 and their eigenvalues always appear in complex conjugate pairs or are +1 \parencite[Theorem 12.10]{Gallier2011}. In odd dimensions there is therefore at least one eigenvalue equal to +1 fixating an axis.

A well-established result from Lie theory \parencite[Theorem~18.1]{Gallier2011} is that each \(n \times n\) rotation matrix \(\mat{R}\) is the image of an \(n \times n\) skew-symmetric\footnote{A skew-symmetric matrix \(\mat{L} \in \R^{n \times n}\) fulfills \(\mat{L} = -\mat{L}^\top\).} matrix \(\mat{L}\) under the matrix-exponential function, \ie \(\mat{R} = \exp(\mat{L})\).
Due to the fact that \(\mat{R}\) is determined by \(\mat{L}\), it is called a \emph{generator} of \(\mat{R}\).

Introduce now an additional parameter \(\theta\), interpreted as an angle, and write \(\mat{R}(\theta) = \exp(\theta \mat{L})\). 
Considering an infinitesimal angle \(d\theta\) and expanding the expression for \(\mat{R}(\theta)\) results in \(R(d\theta) = \mat{I}_n + d\theta \mat{L}\), using as usual that powers of \(d\theta\) with exponent \(\geq 2\) vanish.
Given two unit vectors \(\bm{u}_1\) and \(\bm{u}_2\) orthogonal to each other, we are interested in the rotation matrix of a simple rotation transforming \(\R^n\) by rotating parallel to the subspace spanned by \(\bm{u}_1\) and \(\bm{u}_2\).
We can assume without loss of generality that rotation occurs from \(\bm{u}_1\) towards \(\bm{u}_2\) while keeping the orthogonal complement of the rotating subspace fixed.
It is therefore required that \(\mat{R}(d\theta) \bm{u} = \bm{u}\) for any \(\bm{u}^\top \perp \bm{u}_1, \bm{u}_2\). The infinitesimal linearization \((\mat{I}_n + d\theta \mat{L}) \bm{u} = \bm{u}\) then implies that \(\mat{L} \bm{u} = \bm{0}\) whenever \(\bm{u} \perp \bm{u}_1, \bm{u}_2\).
Furthermore, since we are interested in the rotation from \(\bm{u}_1\) towards \(\bm{u}_2\) we expect that \(\mat{R}(d\theta) \bm{u}_1 = \bm{u}_1 + d\theta \bm{u}_2\) which implies that \(\mat{L} \bm{u}_1 = \bm{u}_2\) should hold, and, analogously, \(\mat{L} \bm{u}_2 = -\bm{u}_1\).
This defines how \(\mat{L}\) acts on \(\R^n\) and \(\mat{L} = \bm{u}_2 \bm{u}_1^\top - \bm{u}_1 \bm{u}_2^\top\) fulfills all requirements.

Exponentiating this matrix results in
\begin{align*}
\mat{L}^2 &=
-(\bm{u}_1 \bm{u}_1^\top + \bm{u}_2 \bm{u}_2^\top), \\
\mat{L}^3 &=
-(\bm{u}_2 \bm{u}_1^\top - \bm{u}_1 \bm{u}_2^\top) = -\mat{L}, \\
\mat{L}^4 &=
\bm{u}_1 \bm{u}_1^\top + \bm{u}_2 \bm{u}_2^\top, \\
\mat{L}^5 &= \mat{L},
\end{align*}
and the matrix exponential therefore evaluates to
\begin{equation*}
\begin{aligned}
\mat{R}(\theta) &= \exp(\theta \mat{L}) \\
&= \mat{I}_n + \sin(\theta) (\bm{u}_2 \bm{u}_1^\top - \bm{u}_1 \bm{u}_2^\top) \\
&+ (\cos(\theta) - 1) (\bm{u}_1 \bm{u}_1^\top + \bm{u}_2 \bm{u}_2^\top).
\end{aligned}
\end{equation*}
We obtain the explicit parametrization of a rotation matrix which rotates \(\bm{u}_1\) towards \(\bm{u}_2\) by angle \(\theta\).

In the context of SVD a data singular vector \(\bm{a}\) corresponding to a signal singular vectors \(\bm{u}\) can always be seen as a rotation of the latter by an angle \(\theta\) in a plane spanned by \(\bm{u}\) and a second unknown unit vector \(\bm{v}\) orthogonal to \(\bm{u}\). This means that
\begin{equation*}
\bm{a} = \mat{R}(\theta) \bm{u}
= \cos(\theta) \bm{u} + \sin(\theta) \bm{v}.
\end{equation*}
Note that while the angle in \cref{lmm:asymp-angles-values} can only be computed from the singular values asymptotically, the result about rotation is non-asymptotic and holds always.
The second vector \(\bm{v}\) can be chosen as the normalized form of \(\bm{a} - \bm{a}^\top \bm{u} \bm{u}\) and \(\theta\) is defined by \(\bm{a}^\top \bm{u} = \cos(\theta)\).


\section{Large-scale Collective Matrix Factorization}%
\label{sec:lscmf}

Let the available views be indexed by \(i \in \mathcal{V} = \{1, \dots, m\}\) and denote the set of observed relations \(\mathcal{I} = \{(i, j) : \mat{Y}_{ij} \text{ is observed}\}\).
Assume that every observed matrix is of the form
\begin{equation}\label{eq:data-int-model}
\mat{Y}_{ij} = \mat{X}_{ij} + \sigma_{ij} \mat{Z}_{ij}
\end{equation}
where \(\mat{Y}_{ij}, \mat{X}_{ij}, \mat{Z}_{ij} \in \R^{p_i \times p_j}\),
the elements of \(\mat{Z}_{ij}\) are independently distributed as described
in \cref{ass:entries-z}, and the variance is \(\sigma_{ij} > 0\).
We assume that these matrices are following the asymptotic model in\cref{eq:asymp-model} and therefore are assumed to grow in size with
\(p_i / p_j \rightarrow \beta_{ij}\) for \(k \rightarrow \infty\). However, to simplify notation, we will drop the index \(k\) in the following.

Note that tensor-like data can be described in this model as well. To do so, an additional index \(\gamma\) can be introduced to indicate layers. In that case, \(\mathcal{I} = \{(i, j, \gamma) : \mat{Y}_{ij, \gamma} \text{ is observed}\}\) and indices in \cref{eq:data-int-model} change accordingly. However, to keep notation light we do not explicitly include \(\gamma\) in the following. All presented results hold regardless also for the tensor-like case.

Assume that each view \(i\) is associated with a \(r_i\)-dimensional subspace \(\mathcal{U}_i\) of \(\R^{p_i}\).
Set \(r := \max_i r_i\), assume that \(r < p_j\) for all \(j\) and denote by \(\mat{U}_i \in \R^{p_i \times r}\) orthogonal matrices that contain a basis for \(\mathcal{U}_i\). If \(r_i < r\), then \(\mat{U}_i\) contains \(r - r_i\) orthogonal vectors from the orthogonal complement of \(\mathcal{U}_i\).

To facilitate data integration, assume that the signal matrices \(\mat{X}_{ij}\) are low-rank and that in addition to \cref{ass:signal-sing-vals} it holds that
\begin{equation}\label{eq:data-int-signal}
\mat{X}_{ij} =
\sum_{l = 1}^r \bm{x}_{ij}^{(l)} \mat{U}_i^{(:, l)} {\mat{U}_j^{(:, l)}}^\top,
\end{equation}
i.e.\@ the left singular vectors only depend on view~\(i\) and the right singular vectors only depend on view~\(j\).  Note that the singular vectors can change role from left to right singular vectors depending on whether they appear in \(\mat{X}_{ij}\) or \(\mat{X}_{ji}\). The elements of \(\bm{x}_{ij}\) are not required to be sorted in descending order, can be positive and negative, and are allowed to be zero. The case \(\bm{x}_{ij} = \bm{0}\) is theoretically allowed but in practice not of interest.

\begin{remark}
  Since some elements of \(\bm{x}_{ij}\) are allowed to be zero, it is likely
  that it will be impossible to recover the entirety of \(\mat{U}_i\) from
  a single individual matrix. One goal of data integration is to obtain
  a more complete picture by combining the information from multiple
  individual data matrices.
\end{remark}

Each data matrix \(\mat{Y}_{ij}\) is allowed to have its own noise variance
\(\sigma_{ij}\). However, to simplify analysis in the following let
\(p_{ij} = \max(p_i, p_j)\) and standardize the matrices to
\begin{equation}\label{eq:data-int-model-norm}
\mat{Y}_{ij} = \mat{X}_{ij} + \frac{1}{\sqrt{p_{ij}}} \mat{Z}_{ij}.
\end{equation}
Note that \(\mat{Y}_{ij}\) and \(\mat{X}_{ij}\) are scaled versions of their
counterparts in \cref{eq:data-int-model}, but for ease of notation the same
symbols will be used as previously.

In practice, this scaling can be approximately achieved by dividing
each entry in the unscaled \(\mat{Y}_{ij}\) by \(\sqrt{p_{ij}} \hat{\sigma}\)
where \(\hat{\sigma}\) is the estimator in \cref{eq:sd-estimator} for
\(\mat{Y}_{ij}\) if \(p_i \leq p_j\) and \(\mat{Y}_{ij}^\top\) otherwise.

\subsection{Joint model for a fixed view}

All data matrices involving a specific view~\(i\) can be collected in a matrix \(\mat{Y}_i\) with view~\(i\) in the rows and all other views, which are observed in a relationship with view~\(i\), in the columns, concatenated in a column-wise fashion.
Transposition of the individual data matrices might be necessary before concatenation if view~\(i\) appears in the columns of a data matrix. Therefore, matrix \(\mat{Y}_i\) will have \(p_i\) rows and \(c_i = \sum_{(i, j) \in \mathcal{I}} p_j + \sum_{(j, i) \in \mathcal{I}} p_j\) columns.
To make it easier to work with these joint matrices, each involved data matrix \(\mat{Y}_{ij}\) or \(\mat{Y}_{ji}^\top\) will be standardized to noise variance 1 first, i.e.\@ \(\sqrt{p_{ij}} \mat{Y}_{ij}\) or \(\sqrt{p_{ij}} \mat{Y}_{ji}^\top\) will be joined together, and the matrix \(\mat{Y}_i\) will be divided by \(\sqrt{d_i}\) where \(d_i = \max(p_i, c_i)\) to ensure that \(\mat{Y}_i\) is of the same form as the matrices in the data integration model \cref{eq:data-int-model-norm}.

Formally, these matrices can be written as
\begin{equation}\label{eq:joint-view-mat}
\begin{alignedat}{2}
  \mat{Y}_i &=& \frac{1}{\sqrt{d_i}} \Big[&
    \sqrt{p_{ij}} \mat{Y}_{ij} \text{ for } (i, j) \in \mathcal{I}
    \ \Big\vert\ %
    \sqrt{p_{ji}} \mat{Y}_{ji}^\top \text{ for } (j, i) \in \mathcal{I}
  \Big] \\
  &=& \frac{1}{\sqrt{d_i}} \Big[&
    \sum_{l = 1}^r \sqrt{p_{ij}} \bm{x}_{ij}^{(l)}
    \mat{U}_i^{(:, l)} {\mat{U}_j^{(:, l)}}^\top
    \text{ for } (i, j) \in \mathcal{I}
    \ \Big\vert\ \\
    & & &
    \sum_{l = 1}^r \sqrt{p_{ji}} \bm{x}_{ji}^{(l)}
    \mat{U}_i^{(:, l)} {\mat{U}_j^{(:, l)}}^\top
    \text{ for } (j, i) \in \mathcal{I}
  \Big] \\
  &+& \frac{1}{\sqrt{d_i}} \Big[&
    \mat{Z}_{ij} \text{ for } (i, j) \in \mathcal{I}
    \ \Big\vert\ %
    \mat{Z}_{ji}^\top \text{ for } (j, i) \in \mathcal{I}
  \Big] \\
  &=& \sum_{l = 1}^r & \bm{x}_i^{(l)} \mat{U}_i^{(:, l)} {\mat{V}_i^{(:, l)}}^\top
  + \frac{1}{\sqrt{d_i}} \mat{Z}_i
\end{alignedat}
\end{equation}
where for \(l \in [r]\)
\begin{equation}\label{eq:joint-signal-singular-value}
\bm{x}_i^{(l)} =
\sqrt{
\frac{1}{d_i}\left(
\sum_{(i, j) \in \mathcal{I}} p_{ij} {\bm{x}_{ij}^{(l)}}^2 +
\sum_{(j, i) \in \mathcal{I}} p_{ji} {\bm{x}_{ji}^{(l)}}^2\right)},
\end{equation}
\(\mat{Z}_i\) is the column-wise concatenation of all \(\mat{Z}_{ij}\) and
\(\mat{Z}_{ji}^\top\), and using row-wise concatenation
\begin{equation*}
\begin{aligned}
\bm{V}_i &=
\begin{bmatrix}
\sqrt{p_{ij}} \bm{U}_j
\Diag\left(\bm{x}_{ij}^{(1)}, \dots, \bm{x}_{ij}^{(r)}\right)
& \text{ for } (i,j) \in \mathcal{I} \\
\sqrt{p_{ji}} \bm{U}_j
\Diag\left(\bm{x}_{ji}^{(1)}, \dots, \bm{x}_{ji}^{(r)}\right)
& \text{ for } (j,i) \in \mathcal{I} \\
\end{bmatrix} \\
&\cdot
\Diag\left(
\sqrt{d_i} \bm{x}_{i}^{(1)}, \dots, \sqrt{d_i} \bm{x}_{i}^{(r)}
\right)^{-1}.
\end{aligned}
\end{equation*}
Straight-forward computation shows that \(\bm{V}_i\) is an orthogonal matrix.
Note how the signal's left-singular vectors \(\bm{U}_i\) are unchanged and appear in the same form as in the original data matrices. Also, the joint matrix \(\mat{Y}_i\) is not simply a concatenation of observed matrices but also takes correct scaling of each matrix into account. This will be important in the following analysis.

\subsection{Estimating a view-specific factor match graph}

Since the joint matrix \(\mat{Y}_i\) follows the same model structure in \cref{eq:asymp-model} as individual data matrices, it is possible to use the matrix denoising in \cref{eq:denoised-signal} to recover the signal \(\hat{\mat{X}}_{ij}\) from each individual data matrix \(\mat{Y}_{ij}\), as well as the signal \(\hat{\mat{X}}_i\) for the joint data matrix \(\mat{Y}_i\). \cref{eq:joint-signal-singular-value} then shows that, signal present in individual matrices propagates to signal present in the joint matrix. \cref{sec:properties-joint-mat} explores different scenarios in which signal is likely to be retained in the joint matrix and describes situations in which weak signal can be boosted or drowned out.

Let \(\mat{A}_i\) and \(\mat{A}_{ij}\) be the left singular vectors of \(\hat{\mat{X}}_i\) and \(\hat{\mat{X}}_{ij}\), respectively. Note that we need to consider right singular vectors \(\mat{B}_{ij}\) in case of \(\hat{\mat{X}}_{ji}\). For simplicity, we will restrict discussion to the left singular vector case. Both are approximations of \(\mat{U}_i\), or rather, most likely of a subset of columns thereof. By comparing the columns of \(\mat{A}_i\) and \(\mat{A}_{ij}\) it is possible to determine which factors appear in both matrices. By doing so for each \(\mat{X}_{ij}\) involved in \(\mat{Y}_i\) a factor match graph can be constructed from the perspective of view \(i\). Formally, we generate a hypergraph \(H_i = (F_i, C_i)\) such that \(f_{j,l} \in F_i\) if column \(\mat{A}_{ij}^{(:, l)}\) matches with column \(k\) in \(\mat{A}_i\). In addition, we add \(f_{j,l}\) to hyperedge \(c_k \in C_i\). In words, each left singular vector from an individual matrix matching a specific left singular vector in the joint matrix will end up in the same hyperedge.

The simplicity of this procedure is appealing, however, it glosses over how to actually perform matching of factors. A possible approach is described in the next section.

\subsection{Geometric factor matching}

The goal of our approach is to determine whether two empirically determined factors from two different denoising estimates represent the same underlying factor. If so, they should be matched and considered to be one factor.
In other words, given empirical unit singular vectors \(\bm{a}_1\) and \(\bm{a}_2\) we would like to determine whether the direction of variance they describe is approximately the same. This can be done by determining whether the directions described by \(\bm{a}_1\) and \(\bm{a}_2\) are approximately parallel, i.e. \(\bm{a}_1^\top \bm{a}_2 \approx \pm 1\). Matching can then be performed by observing the scalar products between factors. Instead of relying on some arbitrary ad-hoc criterion, such as \(\abs{\bm{a}_1^\top \bm{a}_2} \geq 0.75\), the asymptotic geometry of singular vectors described in \Cref{ssec:asymptotic-geometry-singular-vectors} will be utilized to derive theoretically motivated cut-off values.

In the following, the scalar product between two vectors \(\bm{a}_1\) and \(\bm{a}_2\), both rotations of \(\bm{u}\) by angles \(\theta_1\) and \(\theta_2\), respectively, will be considered. As shown in \cref{ssec:high-dim-rotations}, there exist unit vectors \(\bm{v}_1\) and \(\bm{v}_2\) orthogonal to \(\bm{u}\) such that
\begin{equation}\label{eq:rot-same-u}
\bm{a}_i = \cos(\theta_i) \bm{u} + \sin(\theta_i) \bm{v}_i\quad\text{for } i \in \{1, 2\}.
\end{equation}

Note that without loss of generality \(\theta_i \in [0, \pi/2]\). If \(\theta_i \in [\pi, 2\pi]\) then choosing \(\overline{\bm{v}}_i = -\bm{v}_i\) and \(\overline{\theta}_i = 2\pi - \theta_i\) ensures that \(\overline{\theta}_i \in [0, \pi]\) and \(\cos(\overline{\theta}_i) \bm{u} + \sin(\overline{\theta}_i) \overline{\bm{v}}_i = \cos(\theta_i) \bm{u} + \sin(\theta_i) \bm{v}_i = \bm{a}_i\).
If \(\theta_i \in (\pi/2, \pi]\) then \(-\bm{a}_i\) can be considered instead, since we are only interested in whether directions induced by \(\bm{a}_1\) and \(\bm{a}_2\) are approximately parallel. Set \(\overline{\bm{v}}_i = -\bm{v}_i\) and \(\overline{\theta}_i = \pi - \theta_i\). Then \(\overline{\theta}_i \in [0, \pi/2]\) and \(\cos(\overline{\theta}_i) \bm{u} + \sin(\overline{\theta}_i) \overline{\bm{v}}_i = -(\cos(\theta_i) \bm{u} + \sin(\theta_i) \bm{v}_i) = -\bm{a}_i\).
In both cases, a representation of \(\bm{a}_i\) or \(-\bm{a}_i\) can be found that ensures that \(\theta_i \in [0, \pi/2]\).
%

Using \cref{eq:rot-same-u} it then holds that
\begin{equation*}
\bm{a}_1^\top \bm{a}_2 = \cos(\theta_1) \cos(\theta_2)
+ \sin(\theta_1) \sin(\theta_2) \bm{v}_1^\top \bm{v}_2.
\end{equation*}
Since \(\sin(\theta_1) \sin(\theta_2) \geq 0\) due to \(\theta_i \in [0, \pi/2]\) and \(-1 \leq \bm{v}_1^\top \bm{v}_2 \leq 1\), it follows from standard trigonometric results that
\begin{equation}\label{eq:lower-upper-bound-match}
l(\theta_1, \theta_2) = \cos(\theta_1 + \theta_2)
\leq \bm{a}_1^\top \bm{a}_2
\leq \cos(\theta_1 - \theta_2) = u(\theta_1, \theta_2).
\end{equation}
This provides lower and upper bounds \(l\) and \(u\) that are fulfilled if \(\bm{a}_1\) and \(\bm{a}_2\) originate from the same signal factor. However, as can be seen in \cref{fig:asymp-cos-of-angles}, if signal is weak, then \(\theta_i\) can become large in particular the lower bound \(l\) can become too permissive and the risk for spurious matches increases. To prevent this, a bound to exclude non-matches is investigated next.

If \(\bm{a}_1\) and \(\bm{a}_2\) do not originate from the same signal factor, then there are orthogonal unit vectors \(\bm{u}_1\) and \(\bm{u}_2\), as well as unit vectors \(\bm{v}_i\) orthogonal to \(\bm{u}_i\), respectively, such that
\begin{equation*}
\bm{a}_i = \cos(\theta_i) \bm{u}_i + \sin(\theta_i) \bm{v}_i\quad \text{for } i \in {1, 2}.
\end{equation*}
It then follows that
\begin{equation*}
\bm{a}_1^\top \bm{a}_2 =
\cos(\theta_1) \sin(\theta_2) \bm{u}_1^\top \bm{v}_2
+ \cos(\theta_2) \sin(\theta_1) \bm{v}_1^\top \bm{u}_2
+ \sin(\theta_1) \sin(\theta_2) \bm{v}_1^\top \bm{v}_2.
\end{equation*}
Since all cosine and sine are non-negative due to \(\theta_i \in [0, \pi/2]\) the expression can be bounded using standard trigonometric results such that
\begin{equation}\label{eq:lower-upper-bound-non-match}
\begin{aligned}
L(\theta_1, \theta_2) &= -\left(\sin(\theta_1 + \theta_2) + \sin(\theta_1)\sin(\theta_2)\right) \leq \bm{a}_1^\top \bm{a}_2 \\
&\leq \sin(\theta_1 + \theta_2) + \sin(\theta_1)\sin(\theta_2) = U(\theta_1, \theta_2).
\end{aligned}
\end{equation}
This provides lower and upper bounds \(L\) and \(U\) that are fulfilled if \(\bm{a}_1\) and \(\bm{a}_2\) originate from two different signal factors.

It is reasonable to expect that for matching vectors
\begin{equation*}
\bm{a}_1^\top \bm{a}_2 \geq l(\theta_1, \theta_2)
\quad\text{as well as}\quad
\bm{a}_1^\top \bm{a}_2 \geq U(\theta_1, \theta_2).
\end{equation*}
The bounds can therefore be used to make an informed guess on whether or not the two vectors originate from the same signal. The upper bound in \cref{eq:lower-upper-bound-match} could be used as well. However, we have seen little practical usefulness as it is typically close to one and the risk is to exclude actual matches due to noise in the data. 

\begin{figure}[ht]
\centering
\includegraphics[width=\textwidth]{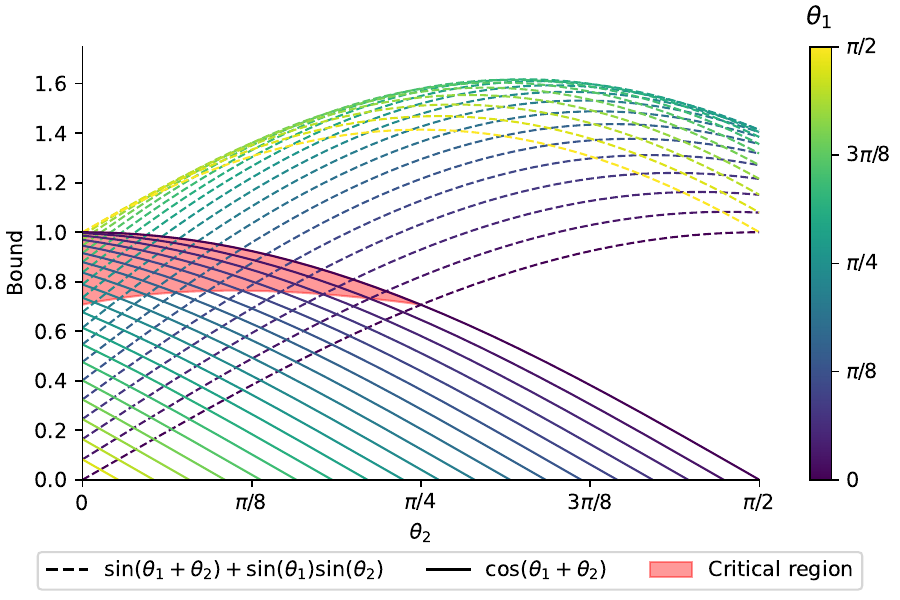}
\caption{Matches are only possible if the lower bound for the matching case in \cref{eq:lower-upper-bound-match} (solid lines) is greater or equal to the upper bound for the non-matching case in \cref{eq:lower-upper-bound-non-match} (dashed lines). Each line represents the behaviour of the two bounds for a fixed angle \(\theta_1\) (denoted by color) across all possible \(\theta_2\). Note that the formulas of the bounds are symmetric in \(\theta_1\) and \(\theta_2\) and shown relationships also hold for reversed roles. The red shaded area indicates the critical region within which matching can happen.}
\label{fig:matching-bounds}
\end{figure}

In addition, we find it reasonable to assume that \(U(\theta_1, \theta_2) \leq l(\theta_1, \theta_2)\) should hold such that decision regions for non-matches and matches do not overlap. By adding this extra assumption, we naturally introduce an upper limit of 45 degrees on how far \(\theta_1\) and \(\theta_2\) can be apart from each other.
\Cref{fig:matching-bounds} shows the relationship between the lower bound in \cref{eq:lower-upper-bound-match} and the upper bound in \cref{eq:lower-upper-bound-non-match}. For two empirical factors \(\bm{a}_1\) and \(\bm{a}_2\) to be considered matching the acceptable bound on their scalar product depends on \(\theta_1\) and \(\theta_2\).

There is, however, one additional complicating factor. In practice, \(\theta_i\) is unknown and needs to be estimated.
\cref{lmm:asymp-angles-values} shows that asymptotically \(\abs{\bm{a}_i^\top \bm{u}} \rightarrow \abs{\cos(\theta_i)}\), where \(\abs{\cos(\theta_i)}\) is asymptotically related to the signal singular value \(x_i\) as described in \cref{eq:asymp-left-angles,eq:asymp-right-angles}. Since \(\bm{u}\) is  inaccessible, an estimate of the signal singular value \(x_i\) is necessary to compute \(\theta_i\).
The relationship between signal singular values and data singular values for \(x_i > \beta^{1/4}\) in \cref{lmm:asymp-angles-values} can be inverted at the asymptotic limit, resulting in
\begin{equation}\label{eq:approx-signal-singular-value}
%
%
%
x_i \approx \sqrt{\frac{1}{2} \left(
  y_i^2 - \beta - 1 + \sqrt{(y_i^2 - \beta - 1)^2 - 4 \beta}
\right)}
\end{equation}
for \(y_i > 1 + \sqrt{\beta}\).
Using this asymptotic estimate for \(x_i\), the value of 
\(\abs{\cos(\theta_i)}\) can be estimated. From this we can retrieve an angle \(\theta_i \in [0, \pi / 2]\) which is sufficient as we argued above.

We are therefore able to derive cut-off values for \(\abs{\bm{a}_1^\top \bm{a}_2}\) that are motivated by the asymptotic geometry of singular vectors. As described in the previous sections, view-specific factor match graphs can then be formed using this matching technique. It remains to aggregate information stored in individual factor match graphs \(H_i\) into a final estimate \(H\) which describes the relationships among factors across all input data matrices.


\subsection{Hypergraph Merging}%
\label{ssec:hypergraph-merging}



The key idea behind the merging algorithm is that each data matrix is always involved in exactly two factor match graphs, since each data matrix is associated with two different views. Therefore, factor nodes corresponding to the same factor will occur among nodes in two different hypergraphs and by merging the hyperedges they are contained withing, the two factor match graphs can be merged into one.

For example, matrix \(\mat{Y}_{12}\) contributes to \(H_1\) and \(H_2\). Let \(f_{2,l} \in F_1\) and \(f_{1,k} \in F_2\) be factors associated with \(\hat{\mat{X}}_{12}\) during factor matching. If they correspond to the same factor in \(\hat{\mat{X}}_{12}\), then this means that \(c_l \in C_1\) and \(c_k \in C_2\) represent the same underlying (possibly shared) factor in the true signal. By merging hyperedges and moving missing factors from \(H_2\) into \(H_1\), an updated hypergraph \(\tilde{H}\) can be formed, that contains the structure of both \(H_1\) and \(H_2\). By iterating through all view-specific factor graphs, a final estimate describing the overall structure can be found.



Denote by \(\mathcal{H}\) the set of view-specific hypergraphs \(H_i\) obtained during the matching step. Merging can then be performed with the following algorithm.
\begin{enumerate}[noitemsep]
\item As long as \(\abs{\mathcal{H}} > 1\), choose two arbitrary hypergraphs \(H_1\) and \(H_2\) from \(\mathcal{H}\), thereby removing them from \(\mathcal{H}\).
Denote the maximal number of factors per view in \(H_i\) as \(k_i\).
\item Find all nodes \(\mathcal{F}\) that appear in both graphs and iterate through them. For node \(f\) in \(\mathcal{F}\):
\begin{enumerate}
\item Check that \(f\) is still present in \(H_2\), since it may have been removed during a previous iteration. If not, continue to the next node.
\item Otherwise, find the incident edge \(c_1\) in \(H_1\) and \(c_2\) in \(H_2\) that contain \(f\). Construction of the view-specific hypergraphs ensures that a node only appears in one hyperedge.
\item Add to \(H_1\) all nodes in \(c_2\) that do not exist in \(H_1\) and add to \(c_1\) all nodes in \(c_2\) not in \(c_1\).
\item Find all other hyperedges in \(H_1\) that \(c_2\) overlaps with (contains shared nodes with) other than \(c_1\). Combine those edges with \(c_1\) as in the previous step.
\item Remove \(c_2\) and all nodes in \(c_2\) from \(H_2\). Nodes other than \(f\) might be removed in this step which requires the check at the beginning of the inner loop.
\end{enumerate}
\item All remaining nodes in \(H_2\) that are not shared between \(H_1\) and \(H_2\) are added to \(H_1\). All remaining edges in \(H_2\) are iterated over and added as new edges to \(H_1\).
\item The updated hypergraph \(H_1\) is added back to \(\mathcal{H}\).
\item Continue with Step 1.
\end{enumerate}

To summarize, lsCMF forms joint matrices from the perspective of each view \(i \in \mathcal{V}\), denoises both joint and individual matrices, performs geometric factor matching to obtain view-specific factor match graphs \(H_i\), and finally merges them into a single factor match graph \(H\).

\section{Implementation details}

Performing lsCMF estimation requires to compute all singular values of all individual data matrices \(\mat{Y}_{ij}\) and all joint matrices \(\mat{Y}_i\) to perform matrix denoising and estimate their rank \(r_{ij}\) and \(r_i\), respectively. Note that no singular vectors are estimated in this first step to improve performance.
Once the rank is known, a truncated SVD can be computed for each individual and each joint matrix, restricted to \(r_i\) or \(r_{ij}\), respectively. Typically, these ranks are small compared to matrix dimensions and therefore this step reduces computation time significantly.

Final estimates for \(\mat{V}_i\) and \(\bm{x}_{ij}\) can be extracted using \(H\). To do so, hyperedges \(c_l\) in \(H\) are used to perform reconstruction. If there are \(r\) hyperedges in total, then the final estimate for \(\mat{V}_i\) will be a \(p_i \times r\) matrix.
Hyperedge \(c_l\) then leads to factor \(l\) in \(\bm{V}_i\) if it contains a factor node which is associated with a data matrix that contains view \(i\). Singular vectors are then obtained from the empirical singular vectors \(\mat{A}_i\) which are computed during denoising of \(\mat{Y}_i\). Taking singular vectors for view \(i\) only from \(\mat{A}_i\) ensures that all columns in \(\mat{V}_i\) are orthogonal. It is possible that there are columns in \(\mat{V}_i\) that are not used since there are no associated factor nodes. These can be set arbitrarily to vector from the orthogonal complement of the already filled in columns. In the current implementation, these columns are simply left empty since they are not used and are essentially arbitrary. 

An active factor in matrix \(\mat{X}_{ij}\) should always be detectable from both directions, i.e., it should appear during matching for views \(i\) as well as \(j\). However, due to noise and the phenomena described in the appendix, it can happen that a factor is only found from one direction. Assume factor \(l\) described by hyperedge \(c_l\) is found in view \(i\) but not view \(j\) and that it corresponds to factor \(k\) in \(\hat{\mat{X}}_{ij}\), i.e. it was found while matching \(\mat{A}_{ij}^{(:, k)}\) with some column in \(\mat{A}_i\). The factor vector for view \(i\) is taken from the empirical singular vectors in \(\mat{A}_i\) as described above. For view \(j\) no singular vector in \(\mat{A}_j\) was found to correspond to the factor. Instead we use the empirical right singular vector \(\mat{B}_{ij}^{(:, l)}\) (or left singular vector \(\mat{A}_{ji}^{(:, l)}\) if \(j\) is the view in the rows). Note that this may lead to columns in \(\mat{V}_i\) not strictly being orthogonal, however, due to the geometry of singular vectors described earlier they should be close to orthogonal.

Singular values \(\bm{x}_{ij}\) are taken equal to the shrunk singular values obtained during denoising in \cref{eq:denoised-signal}. They are arranged in order to line up with the corresponding factors in \(\mat{V}_i\) and \(\mat{V}_j\) and are possibly sign corrected. The sign of \(\bm{x}_{ij}^{(l)}\) is negated if \(\mat{V}_i^{(:, l)}\) is anti-parallel to the corresponding column in \(\mat{A}_{ij}\) or if \(\mat{V}_j^{(:, l)}\) is anti-parallel to the corresponding column in \(\mat{B}_{ij}\). No negation is required if both are parallel or both are anti-parallel. lsCMF can therefore produce solutions with negative singular values. Depending on the data layout it is possible to flip signs in factors and singular values to produce solutions with non-negative singular values alone. However, this is not always the case for more complex layouts.


\section{Evaluation}%
\label{sec:evaluation}

To evaluate the performance of lsCMF we explore its performance on simulated data in comparison to other methods.
We focused on two metrics, the elapsed estimation time in seconds as a function of input data size as well as quality of structure estimation by investigating ranks of shared and individual signals estimated.

\begin{figure}[ht]
\centering
\includegraphics{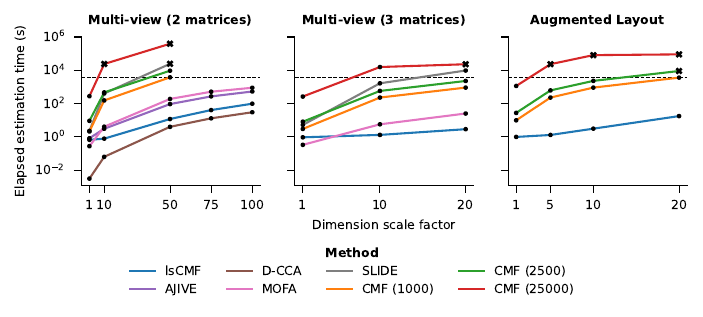}
\caption{Mean elapsed estimation time on a logarithmic scale for three different simulation scenarios. The color of each line indicates the method. Black dots indicate the mean elapsed time in seconds after 25 repetitions. Black crosses indicate the mean elapsed time when less than 25 repetitions could be completed within a 5 day window. The dashed black line marks 1 hour of estimation time.}
\label{fig:sims-elapsed-time}
\end{figure}

We compare against AJIVE \parencite{Feng2018} as implemented in mvlearn \parencite{mvlearn}, D-CCA \parencite{Shu2020}, MOFA \parencite{Argelaguet2018}, SLIDE \parencite{Gaynanova2019}, and CMF \parencite{Klami2014}. However, due to limitations in which matrix layouts can be integrated by these methods, we only used the subset of methods that could be applied to the given data layout without modification. AJIVE, D-CCA, MOFA, and SLIDE only support multi-view data, whereas CMF also supports augmented layouts as well as other more flexible layouts like L-shaped layouts. lsCMF can be applied to any collection of matrices if at least one view in a data matrix overlaps with another the same view in another matrix which is present in the collection.

lsCMF, AJIVE, D-CCA, and SLIDE were used with default settings. MOFA is run with a maximum rank of 10 and without element-wise sparsity in factors, only structure detection. CMF is run with a maximum rank of 10, without bias terms, and the prior parameters of the ARD prior are set to \(\alpha = 1\) and \(\beta = 0.001\) to facilitate factor selection. CMF requires the specification of a maximum number of iterations and does not determine convergence automatically. Clearly, a larger number of iterations increases the overall runtime of the algorithm. We therefore tested to run CMF for 1000, 2500, as well as 25000 iterations and compared the results. MOFA and CMF estimate approximate Bayesian posterior distributions and therefore do not provide binary decisions on which factors are included or excluded. It was therefore necessary to perform post-hoc thresholding of factor scales to decide which factors are part of a data matrices signal. To do so, histograms of factor scales across data matrices and simulation repetitions were investigated to find a suitable cut-off. This would be more difficult in data analysis since thresholds are data dependent. We chose a threshold of 2 for MOFA and CMF in Scenario 1, a threshold of 0.15 for MOFA and CMF in Scenario 2, and a threshold of 1 for CMF in Scenario 3.

Data is simulated following the model in \cref{eq:asymp-model-noise} where \(\sigma\) was chosen such that \(\norm{\mat{X}_{ij}}_F / \sqrt{\sigma^2 p_i p_j} = \mathrm{SNR}\), the signal-to-noise ratio. A normal distribution was assumed for the entries in \(\mat{Z}_{ij}\). Orthogonal singular vectors \(\mat{U}_i \in \R^{p_i \times r}\)are simulated by simulating entry-wise from a standard normal distribution and applying QR factorization. The first \(r\) columns of the \(\mat{Q}\) matrix are then used for \(\mat{U}_i\).

Three simulation setups with increasingly complex matrix layouts were investigated. To evaluate the impact of input size we scaled view dimensions \(p_i\) by a sequence of dimension scale factors.
Simulations were repeated independently 25 times for each combination of simulation scenario and dimension scale factor.
\begin{enumerate}
\item \textbf{Multi-view layout (Two matrices)}: A setup with \(\mathcal{V} = \{1, 2, 3\}\), \(\mathcal{I} = \{(1, 2), (1, 3)\}\), \(p_1 = 100\), \(p_2 = 25\), \(p_3 = 25\),
\begin{equation*}
\begin{array}{rcllll}
\bm{x}_{12} &=& (6, & 7,   & 0, & 8), \\
\bm{x}_{13} &=& (5, & 5.5, & 6, & 0), \\
\end{array}
\end{equation*}
and \(\mathrm{SNR} = 1\). The signals contain shared as well as individual components.
Dimensions \(p_i\) were scaled by 1, 10, and 50 for all methods and even by 75 and 100 for all methods except SLIDE and CMF.
\item \textbf{Multi-view layout (Three matrices)}: A setup with \(\mathcal{V} = \{1, 2, 3, 4\}\), \(\mathcal{I} = \{(1, 2), (1, 3), (1, 4)\}\), \(p_1 = 100\), \(p_i = 25\) for \(i = 2, 3, 4\),
\begin{equation*}
\begin{array}{rclllllll}
\bm{x}_{12} &=& (1.5, & 1.3, & 0.9, & 0.6, & 0,   & 0,   & 0), \\
\bm{x}_{13} &=& (1.5, & 1.3, & 0,   & 0,   & 0.8, & 0.5, & 0), \\
\bm{x}_{14} &=& (1.5, & 1.3, & 1.0, & 0,   & 0,   & 0,   & 0.7), \\
\end{array}
\end{equation*}
and \(\mathrm{SNR} = 1\). The signals contain globally as well as some partially shared, and individual signal.  Dimensions \(p_i\) were scaled by 1, 10, 20.
\item \textbf{Augmented layout}: A setup with \(\mathcal{V} = \{1, 2, 3\}\), \(\mathcal{I} = \{(1, 2), (1, 3), (2, 3)\}\), \(p_i = 100\) for \(i = 1, 2, 3\), 
\begin{equation*}
\begin{array}{rcllllll}
\bm{x}_{12} &=& (0,   & 3.5, & 2.5, & 0,   & 1.9, & 0), \\
\bm{x}_{13} &=& (4.9, & 3.5, & 2.5, & 0,   & 0,   & 2.2), \\
\bm{x}_{23} &=& (4.9, & 3.5, & 0,   & 2.5, & 0,   & 0), \\
\end{array}
\end{equation*}
and \(\mathrm{SNR} = 1\). Globally and partially shared, as well as individual signal is present. Dimensions \(p_i\) were scaled by 1, 5, 10, 20.
\end{enumerate}

\begin{figure}[ht]
\centering
\includegraphics{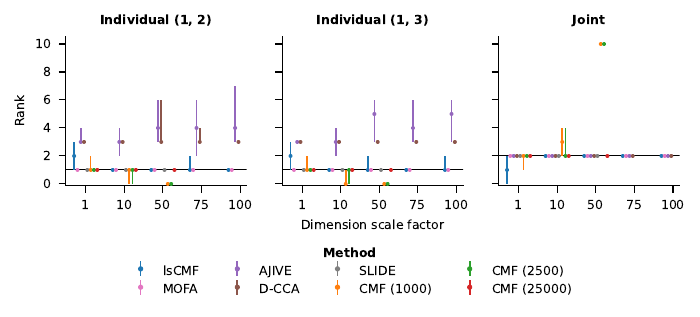}
\caption{Estimated ranks for the two-matrix multi-view simulation. Colors indicate different methods. Dots indicate the median across 25 repetitions (or less if not all 25 repetitions could be obtained in a 5 day window). The bars indicate minimum and maximum estimated rank. Black lines show the groundtruth ranks used during simulation. Note that D-CCA estimates overall individual rank per data matrix and then finds a joint subspace between the individual spaces. Groundtruth for D-CCA is therefore 3 in both individual signals and 2 for the joint signal.}
\label{fig:multiview-2matrices-ranks}
\end{figure}

\begin{figure}[ht]
\centering
\includegraphics{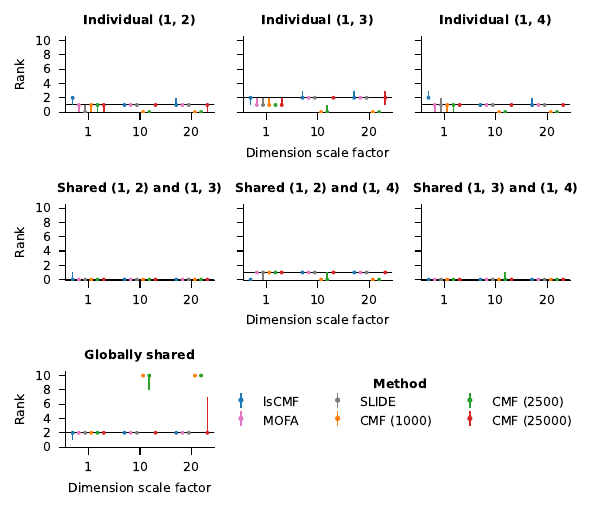}
\caption{Estimated ranks for the three-matrix multi-view simulation. See description of \cref{fig:multiview-2matrices-ranks}. In addition to individual and globally shared factors, there are additional partially shared factors present in this simulation.}
\label{fig:multiview-3matrices-ranks}
\end{figure}

\begin{figure}[ht]
\centering
\includegraphics{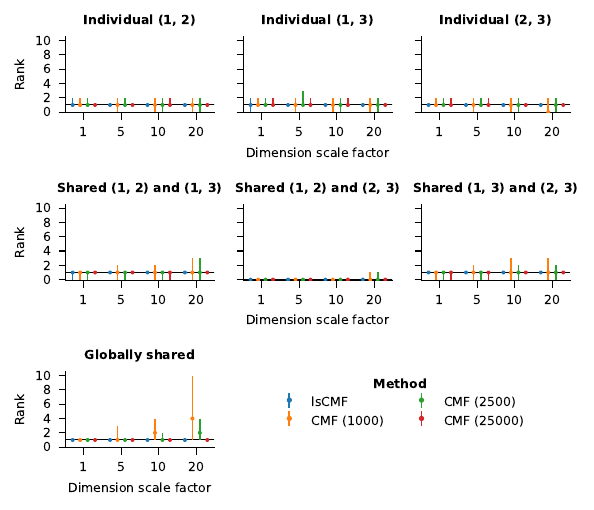}
\caption{Estimated ranks for the augmented simulation. See description of \cref{fig:multiview-3matrices-ranks}.}
\label{fig:augmented-ranks}
\end{figure}

For comparability of runtime, each method was limited to run on a single core of a Intel Xeon Gold 6130 processor and given free access to all available 96GB of RAM. 
Mean elapsed runtime in seconds for each simulation scenario and method is collected in \cref{fig:sims-elapsed-time}. lsCMF performs very well in all scenarios especially for increasing data set size. The only method consistently faster than lsCMF is D-CCA in Scenario 1. However, D-CCA is only applicable to two matrix scenarios. The elapsed estimation time for CMF, naturally, depended on the number of iterations with longer times necessary for higher iteration counts. Note that in some scenarios CMF and SLIDE did not manage to perform estimation on all 25 repeated datasets within a 5 day window.
For Scenario 1 it was decided to increase the dimension scale factor to 75 and 100. However, only lsCMF, D-CCA, MOFA, and AJIVE were run on these large datasets since SLIDE and CMF, even at the lowest number of iterations, already needed a long time at dimension scale factor 50. Trends for estimation time were surprisingly consistent across dimension scale factors. Note how MOFA is faster than lsCMF in both multi-view layouts for low dimensions, however, lsCMF is clearly faster when input size increases.

In addition to elapsed estimation time, separation of ranks into individual, partially shared and globally shared components was investigated. Results for Scenarios 1, 2, and 3 are shown in \cref{fig:multiview-2matrices-ranks,fig:multiview-3matrices-ranks,fig:augmented-ranks}, respectively.
In the two-matrix multi-view scenario, most methods perform well.
lsCMF is based on asymptotic results and therefore profits from increased data size. In \cref{fig:multiview-2matrices-ranks}, it can be seen how lsCMF overestimates individual rank and underestimates shared rank in low dimensions. However, correct ranks are estimated when dimensions increase. A similar effect can be seen in \cref{fig:multiview-3matrices-ranks} for individual ranks and shared ranks between matrices (1, 2) and (1, 4).
AJIVE seems to be having trouble to correctly estimate the individual ranks, but estimates the number of shared components correctly.
Note that D-CCA has slightly different model assumptions than the other methods in this comparison. It estimates subspaces of relevant variation for each data matrix individually and then finds directions of common variation in these subspaces. However, signal is not split into orthogonally complementary subspaces like for lsCMF. The correct groundtruth for D-CCA is therefore 3 for individual ranks and 2 for joint ranks.

An issue with CMF's iteration dependence is apparent in all three simulation scenarios when input data dimensions increase. CMF run for 1000 or 2500 iterations often exhibits a phenomenon where globally shared rank increases to 10 (the maximum allowed rank) and all other individual or partially shared ranks drop to zero. However, when compared to CMF run for 25000 iterations, the later often does find the correct ranks. Investigating these runs shows that CMF must not have converged for larger data sizes and lower maximum iterations, especially since these effects are not apparent for smaller data sizes. However, \cref{fig:sims-elapsed-time} shows how expensive runs for CMF with high iteration count can become. Using CMF is therefore a trade-off between runtime and estimation accuracy.

When investigating the augmented scenario in \cref{fig:augmented-ranks}, lsCMF performs well and estimates ranks correctly throughout all 25 repetitions for dimension scale factor 5 and above.


\section{Discussion}%
\label{sec:discussion}

In this paper, we present a novel way of framing data integration problems as a graphical model and propose a new data integration method focused on scalability. \emph{Large-scale Collective Matrix Factorization (lsCMF)} has been shown, through simulation studies, to be capable of integrating multiple input matrices each as large as 10000 \(\times\) 2500 in seconds and performs comparable to established methods in terms of component structure estimation.
Due to the use of asymptotic theory in the construction of lsCMF, it sometimes performs worse than existing methods for small data sets. However, structure estimation performance is reliable and comparable to existing methods for larger input data sizes.

Note how lsCMF does not require tuning parameters and even preprocessing, at least scaling, is automatically performed during matrix denoising. This works well under the assumptions given in \cref{ssec:model-assumptions}. However, it remains to be investigated how these properties generalize to other noise distributions.

The geometric factor matching approach together with factor match graph merging is a fast and scalable example of a multi-view integration method. The graphical framework for data integration presented in this paper could also be applied to other multi-view integration methods that support detection of shared (global and partial) and individual components. Factor match graphs could then be constructed from those results and merged as described in \cref{ssec:hypergraph-merging}. A motivation for doing so could be to handle other data distributions such as count or binary data. However, this is likely at the cost of estimation speed.

Using SVD to compute the signal matrix requires a dense input matrix. lsCMF therefore does not accommodate integration with missing values in data matrices. If some data sources have missing values and their number is not too numerous, one possibility is to use a matrix imputation algorithm \parencite[e.g.,][]{Mazumder2010,Hastie2015} as preprocessing before data integration.
However, by denoising matrices individually instead of integratively poses a risk that noise will be underestimated during integration. This in turn could lead to matching failures since empirical singular vectors are estimated with false high accuracy. The impact of imputation prior to integration remains to be investigated.
Another approach to consider for future research is to replace the denoising approach \parencite{Gavish2017} with a similar approach supporting missing values \parencite[e.g.,][]{Leeb2021}.



\section*{Acknowledgments}

The computations were enabled by resources provided by the National Academic Infrastructure for Supercomputing in Sweden (NAISS) at National Supercomputer Centre (NSC), Linköping University, partially funded by the Swedish Research Council through grant agreement no. 2022-06725.


\printbibliography

\appendix{
\section*{Appendix}

\begin{figure}[ht]
\includegraphics[width=\textwidth]{%
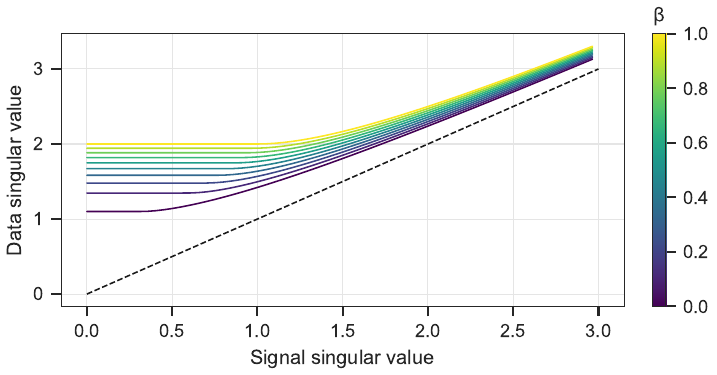}
\caption{Lines show the asymptotic empirical singular value
\(\bm{y}_\infty^{(i)}\) in \cref{eq:asymp-singular-values} dependent on the
corresponding signal singular value \(\bm{x}^{(i)}\) and the asymptotic ratio
\(\beta\) indicated by color.}%
\label{fig:asymp-singular-values}
\end{figure}

\begin{figure}[ht]
  \includegraphics[width=\textwidth]{%
  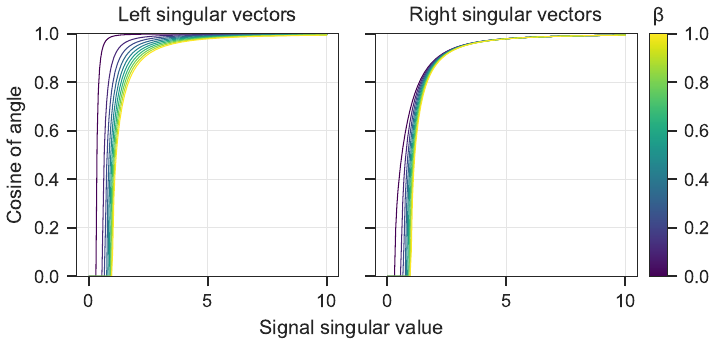}
  \caption{
  Lines show the asymptotic cosine of the angle between the left resp. right
  empirical singular vectors in \(\bm{A}_k^{(:, i)}\) resp.
  \(\bm{B}_k^{(:, i)}\) and the signal left resp. right singular vectors
  in \(\bm{U}_k^{(:, i)}\) resp. \(\bm{V}_k^{(:, i)}\). These depend on
  the corresponding signal singular value \(\bm{x}^{(i)}\) and the asymptotic
  ratio (indicated by color) as shown in
  \cref{eq:asymp-left-angles,eq:asymp-right-angles}.}%
  \label{fig:asymp-cos-of-angles}
\end{figure}

\section{Properties of the joint matrix}%
\label{sec:properties-joint-mat}

It holds that \(1 \geq p_{ij} / d_i\) for all involved \(j\). Any involved
signal singular value \(\bm{x}_{ij}^{(l)}\) therefore either contributes
with its original magnitude or is damped by the factor \(p_{ij} / d_i\).
However, if enough matrices are involved, it is still possible that the
joint signal singular values are larger than the individual ones involved.

In the following set
\begin{equation*}
n_i = \sum_{(i, j) \in \mathcal{I}} 1 + \sum_{(j, i) \in \mathcal{I}} 1
\end{equation*}
to be the number of matrices involving the \(i\)-th view.
To say how \(\bm{x}_i^{(l)}\) behaves in relation to the involved
\(\bm{x}_{ij}^{(l)}\) multiple cases have to be considered. Note that in the
following only \(p_{ij}\) is considered, but all cases also consider
\(i\) to be the second index, \ie \(p_{ji}\).
\begin{enumerate}
\item If \(p_{ij} = p_i\) for all involved \(j\) and even \(d_i = p_i\)
then \(\bm{x}_i^{(l)}\) is just the \(\ell_2\) norm of the involved
\(\bm{x}_{ij}^{(l)}\) seen as a vector. This implies that
\begin{equation*}
\sqrt{n_i} \min_j \bm{x}_{ij}^{(l)}
\leq \bm{x}_i^{(l)}
\leq \sqrt{n_i} \max_j \bm{x}_{ij}^{(l)}
\end{equation*}
where the minimum and maximum are taken over all involved \(j\).
\item If \(p_{ij} = p_i\) for all involved \(j\) but \(d_i = c_i \geq p_i\)
then
\begin{equation*}
\sqrt{\frac{n_i p_i}{c_i}} \min_j \bm{x}_{ij}^{(l)}
\leq \bm{x}_i^{(l)}
\leq \sqrt{\frac{n_i p_i}{c_i}} \max_j \bm{x}_{ij}^{(l)}
\end{equation*}
where \(j\) is taken over all involved \(j\). Note that
\begin{equation*}
\frac{n_i p_i}{c_i} \geq \frac{n_i p_i}{n_i \max_j p_j} \geq 1
\end{equation*}
since \(p_{ij} = p_i\) for all involved \(j\) and where the maximum is
taken over all involved \(j\).
\item If \(p_{ij} = p_j\) for all involved \(j\) then \(d_i = c_i\)
and it is easy to see from the definition of \(c_i\), that
\(\bm{x}_{i}^{(l)}\) is then the weighted average of all involved
\(\bm{x}_{ij}^{(l)}\), with weights equal to \(p_j\). It then holds that
\begin{equation*}
\min_j \bm{x}_{ij}^{(l)} \leq \bm{x}_i^{(l)} \leq \max_j \bm{x}_{ij}^{(l)}
\end{equation*}
where \(j\) is taken over all involved \(j\).
\item If there are \(j_1\) and \(j_2\) such that \(p_{ij_1} = p_{j_1}\)
(and therefore \(d_i = c_i\)) but also \(p_{ij_2} = p_i\), then
\begin{equation*}
\gamma_i :=
\sum_{(i, j) \in \mathcal{I}} p_{ij} + \sum_{(j, i) \in \mathcal{I}} p_{ji}
\geq
\sum_{(i, j) \in \mathcal{I}} p_j + \sum_{(j, i) \in \mathcal{I}} p_j
= c_i.
\end{equation*}
It therefore holds that
\begin{equation*}
\sqrt{\frac{\gamma_i}{c_i}} \min_j \bm{x}_{ij}^{(l)}
\leq \bm{x}_i^{(l)}
\leq \sqrt{\frac{\gamma_i}{c_i}} \max_j \bm{x}_{ij}^{(l)}
\end{equation*}
\end{enumerate}
The four cases cover all possible scenarios. If \(n_i > 1\), it is never
possible for \(\bm{x}_i^{(l)}\) to be smaller than the smallest involved
\(\bm{x}_{ij}^{(l)}\). Also, if \(n_i > 1\), then Cases 1 and 4
actually guarantee that \(\bm{x}_i^{(l)}\) will be larger than the
smallest involved \(\bm{x}_{ij}^{(l)}\). Same holds in Case 2 as long
as \(p_i > \max_j p_j\). The cases above also describe upper limits,
which, just as for the minimum cases, can be larger than the largest
involved \(\bm{x}_{ij}^{(l)}\).

By combining the individual matrices, the joint matrix will often have a
different asymptotic ratio compared to the asymptotic ratios of the
individual matrices. Note that even if the asymptotic ratio increases, and
therefore bias increases as well, \cfr
\cref{fig:asymp-singular-values,fig:asymp-cos-of-angles}, the inequalities
above show that signals not discoverable in the individual matrices can
be discoverable in the joint matrix, at least if they appear in multiple
individual matrices. \cref{ex:signal-amplification} below illustrates
this situation. In contrast, it is possible that forming the joint matrix
makes weak signals undiscoverable as is explore in
\cref{ex:signal-suppression}.

\begin{example}\label{ex:signal-amplification}
Assume there is a signal associated with singular vector \(l\) in view \(i\)
and singular values \(0 < \bm{x}_{ij}^{(l)} < 1 + \sqrt{\beta_{ij}}\) for all
individual matrices involving view \(i\). Notice that all are larger than
zero and therefore \(0 < \min_j \bm{x}_{ij}^{(l)}\).

Since the individual signal singular values are below the threshold of
\(1 + \sqrt{\beta_{ij}}\) it is unlikely that the signal is discovered
in any individual data matrix. However, if we \eg in Case 1 above,
then \(\sqrt{n_i} \min_j \bm{x}_{ij}^{(l)} \leq \bm{x}_i^{(l)}\) which
might be enough to get \(\bm{x}_i^{(l)}\) over the threshold and the signal
therefore discoverable. This is possible if
\((1 + \sqrt{\beta_i})/ \sqrt{n_i} \leq \min_j \bm{x}_{ij}^{(l)}\).
Similar results hold for Cases 2 and 4.

To make the example more concrete, assume there are two matrices with
asymptotic ratio \(0.1\) such that view \(i\) is in the rows and there are
less columns than rows. The theoretical threshold for detectable signals
is then \(1 + \sqrt{0.1} \approx 1.32\) for each matrix individually.
Forming the joint matrix as described above will lead to a matrix following
Case 1 and with asymptotic ratio \(0.2\). Signals that appear in both
matrices with singular values
\(\geq (1 + \sqrt{0.2}) / \sqrt{2} \approx 1.02\) are now detectable.

This example shows that combining information that is present in multiple
data sources can be amplified in the joint matrix.
\end{example}

\begin{example}\label{ex:signal-suppression}
Conversely, weak signals from small data sources can be drowned out by the
addition of larger data sources.

Assume there is a weak signal \(\bm{x}_{ij_1}^{(l)} = 1.25\) in a thin data
matrix with \(p_{j_1} < p_i\) and \(\beta_{ij_1} = 0.05\). The signal is
likely to be detectable in this matrix since
\(1.25 \geq 1 + \sqrt{0.05} \approx 1.22\).
However, assume there is a second data matrix with \(p_{j_2} < p_i\) and
\(\beta_{ij_2} = 0.9\), and \(\bm{x}_{ij_2}^{(l)} < 1.25\). Then
\begin{equation*}
\max_j \bm{x}_{ij}^{(l)} = 1.25
< (1 + \sqrt{0.95}) / \sqrt{2} \approx 1.40
\end{equation*}
which means that the signal will likely not be detectable in the joint matrix.
\end{example}
}

\end{document}